\documentclass{article}
\usepackage[T1]{fontenc}
\usepackage[utf8]{inputenc}
\usepackage{ismir} 
\usepackage{amsmath,cite,url}
\usepackage{amssymb}
\usepackage{graphicx}
\usepackage{wrapfig}
\usepackage{color}
\usepackage{algorithmic}
\usepackage{textcomp}
\usepackage{xcolor}
\usepackage{float}
\usepackage{booktabs}
\usepackage{dblfloatfix}
\graphicspath{{./}{figures/}}

\title{Diff2Mix: Controllable Music Mixing via Diffusion Models and Differentiable Audio Effects}

\oneauthor
  {Yisu Zong  \hspace{1cm}   Jinjie Shi  \hspace{1cm}    Joshua Reiss}
  {Centre for Digital Music, Queen Mary University of London, UK\\\texttt{y.zong@qmul.ac.uk}}

\def\authorname{Y. Zong, J. Shi and J. Reiss}

\begin{document}

\maketitle

\begin{abstract}
Automatic music mixing aims to combine multitrack recordings into a balanced and coherent musical piece. Because the content of different songs and the subjective preferences of mixing engineers jointly shape the final outcome, a practical system should deliver well-balanced mixes while allowing for controllable stylistic variation. However, most existing methods treat automatic mixing and mixing style control as separate tasks, making it difficult for a single system to produce high-quality mixes while remaining editable and style-aware. To address this limitation, this paper presents Diff2Mix, a generative automatic mixing system based on diffusion models and a differentiable mixing console. This system offers two levels of optional user control: a reference audio enables overall production style control, and the differentiable mixing console provides explicit audio effects parameters for interpretability and fine-grained optimization. We demonstrate our system’s competitive performance through both objective and subjective evaluations in terms of mixing quality and control ability. We provide code and audio samples at our project page\footnote{https://zys711.github.io/Diff2Mix}.

\end{abstract}

\section{Introduction}
\begin{figure}[t]\label{overview}
  
  \centering
  \includegraphics[width=\columnwidth]{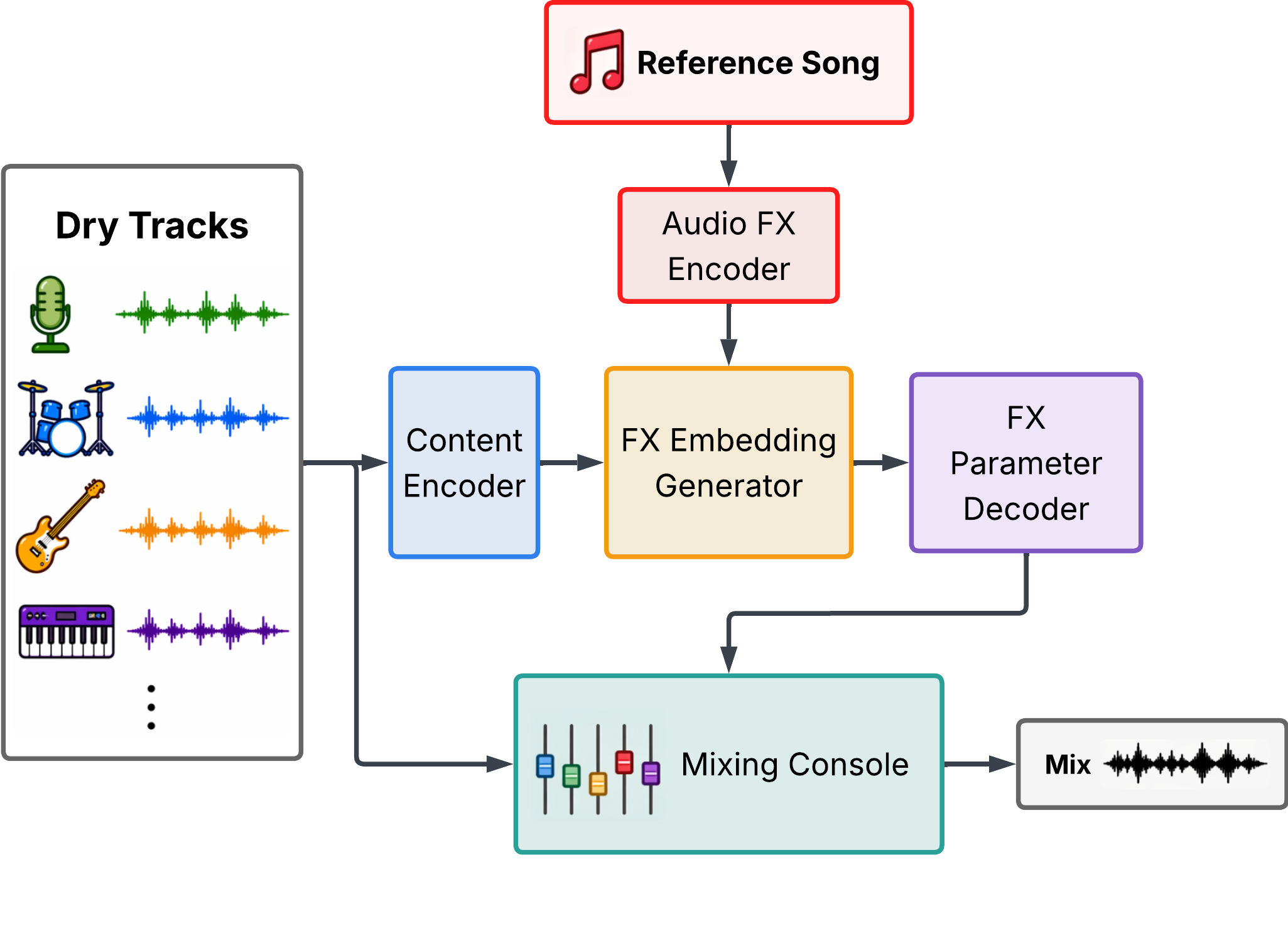}
  \caption{System overview of Diff2Mix.}
  \label{fig:system_overview}
  
\end{figure}

Music mixing is a core stage of music production, where individual tracks are combined into a coherent and expressive final song through the application of audio effects such as equalization, compression, panning, and reverberation. In practice, mixing involves a tightly coupled set of technical and creative decisions. These include balancing audio sources, reducing masking, shaping the dynamics, controlling the spatial impression, and attaining an overall aesthetic goal \cite{case2012mix}. Since these decisions depend both on the musical content and the subjective intent of the mix engineer, producing a high-quality mix remains a difficult and time-consuming task that typically requires substantial expertise. This challenge has motivated growing interest in intelligent music production systems that can automate or assist professional mixing workflows.

Early approaches were dependent on hand-crafted rules derived from engineering practice, offering interpretability but limited flexibility and weak generalization to diverse musical material \cite{de2017ten}. Later methods employed machine learning to learn mixing decisions directly from data. Among these methods, deep learning methods have achieved significant progress. These deep learning approaches can be generally classified into black-box systems \cite{martinez2021deep,marco_a_martinez_ramirez_2022_7316688}, which predict processed audio or full mixes in an end-to-end manner, and parameter estimation systems \cite{steinmetz2021automatic,soumya_sai_vanka_2024_14877399}, which infer control values for audio effects. Black-box models are capable of capturing complex nonlinear mappings; nevertheless, they might generate artifacts or overly averaged results. The applications of Differentiable Digital Signal Processing (DDSP) \cite{engel2019ddsp} in the audio research field have promoted the development of differentiable audio effects \cite{nercessian2020neural,colonel2022direct,steinmetz2021filtered}. This presents a promising solution by incorporating audio effects into end-to-end parameter estimation systems \cite{steinmetz2021automatic,colonel2021reverse}, enabling greater interpretability and control. However, they are commonly restricted to fixed effect chains and differentiable effects. Moreover, these methods typically formulate automatic mixing as a deterministic one-to-one mapping from input tracks to a single target mix. In practice, the same multitrack session can support multiple valid mixes, reflecting different production styles and artistic preferences. Recently, generative model approaches \cite{moliner2026automatic,wu2024diffusion} for automatic mixing have been put forward, which take into account the variety of mixing decisions and have achieved state-of-the-art mixing quality.

A central limitation of many previous approaches is that they lack control over the production style, which is of great significance for collaboration with users. A substantial amount of research on production style transfer aims to capture the reference style and map it onto the target audio, including black-box methods \cite{koo2022end,koo2023music}, parameter estimation methods \cite{steinmetz2022style,soumya_sai_vanka_2024_14877399}, and training-free optimization methods \cite{christian_j_steinmetz_2024_14877423}. With the rapid advancement of large language models and multimodal models, text control could become a novel and promising production style control interface \cite{chu2025text2fx,doh2026llmfxtools}. However, the majority of these methods conduct style control on the mixes rather than the dry tracks. Such systems may struggle to capture the track-specific characteristics and inter-track interactions, and they have not been integrated with automatic mixing systems. Some work \cite{koo2023music,soumya_sai_vanka_2024_14877399} can perform style transfer of multitrack input given a reference, yet these systems only consider style similarity, not overall balance, which limits output mixing quality.

In this work, we propose a diffusion-based automatic multitrack mixing system that controls the overall production style using a reference song. First, we train a track-level audio effects embedding diffusion model conditioned on the musical content of dry tracks and the audio effects features of the reference song. Then, we compare two methods for translating the embedding into audio effects parameters via a differentiable mixing console: directly estimating the parameter using spectral loss, and estimating the parameter combination distribution. We conduct objective and subjective evaluations of both mixing quality and style transfer to demonstrate the effectiveness of our proposed method.  

\section{Method}\label{sec:method}
\begin{figure*}[t]
    \centering
    \includegraphics[width=\textwidth]{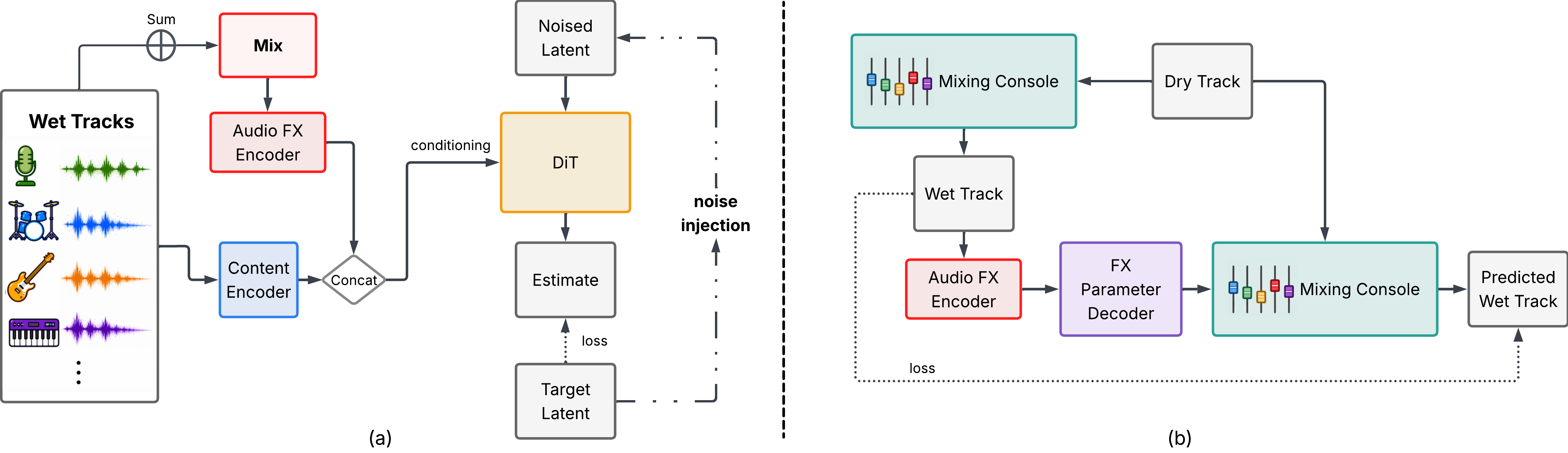}
    \caption{Training pipeline of proposed methods. (a) represents the training of the conditional diffusion model. (b) represents the training of the audio effects parameter decoder.}
    \label{train}
\end{figure*}

The system architecture is shown in Figure \ref{overview}. The first component is a diffusion model conditioned on a reference song's production style, which generates the track-level audio effects embeddings. Subsequently, these embeddings are translated into the audio effects parameters via an audio effects parameter decoder, and these parameters and dry tracks are then passed to a differentiable mixing console to generate the mix.
\subsection{Conditional Audio Effects Embedding Diffusion Model}
Our diffusion model backbone is adapted from MEGAMI \cite{moliner2026automatic}, a diffusion framework operating in the track-level audio effects embedding space. Let \(Z=\{ z_i \}_{i=1}^{n}\) denote the target audio effects embeddings and \(C=\{ c_i \}_{i=1}^{n}\) the content embeddings extracted from the input dry tracks, where \(n\) is the number of tracks. Given a reference song \(y_{ref}\), our aim is to extract a global production-style embedding \(z_{ref}\) and model the conditional distribution \(p_\theta(Z \mid C, z_{ref})\). In line with the EDM formulation \cite{karras2022elucidating}, generation is cast as denoising under a continuous family of Gaussian perturbations rather than a discrete-time forward Markov chain. The forward diffusion process is defined as:

\begin{equation}
\tilde{z}_{i,\sigma} = z_i + \sigma \epsilon_i,\quad
\epsilon_i \sim \mathcal{N}(0, I_D),\quad i = 1, \dots, n
\end{equation}

where \(z_i\) denotes the clean audio effects embedding of the \(i\)-th track, \(\tilde{z}_{i,\sigma}\) is its noised latent at noise level \(\sigma\), \(\epsilon_i\) is Gaussian noise, and \(I_D\) is the \(D \times D\) identity matrix with \(D\) the embedding dimension. The denoiser is trained to recover the clean embedding from the noised input conditioned on the content embeddings \(C\) and the reference style embedding \(z_{ref}\). Equivalently, the network can be interpreted as learning the conditional score:
\begin{equation}
s_{\theta}(\tilde{Z}_{\sigma}, \sigma, C, z_{ref}) \approx \nabla_{\tilde{Z}_{\sigma}} \log p(\tilde{Z}_{\sigma} \mid C, z_{ref})
\end{equation}
The score network is designed to be permutation-equivariant with respect to the track set, so permuting the conditioning tracks leads to the same permutation in the predicted outputs. We implement this property with a Diffusion Transformer (DiT) \cite{peebles2023scalable} that applies self-attention over the noised embedding set \(\tilde{Z}_{\sigma}\) and cross-attention to the conditioning set. In our setting, the reference style embedding \(z_{ref}\) is concatenated to each content embedding \(c_i\) before conditioning, so the network jointly receives track-wise musical content and global production-style information. To maintain correspondence between tracks across the noised embeddings and the conditioning inputs, we append a one-hot index encoding of track position to the representation of each track element. During training, track order is randomly permuted so that the model does not rely on any fixed sequence position as a proxy for instrument identity or musical role. 

Since our objective is to achieve optional reference style control, and it is highly likely that the content of the dry tracks and the reference song will differ, it is necessary to disentangle the control signals of the audio effects style and musical content during training. To support both automatic and reference-guided mixing, we use classifier-free guidance (CFG) \cite{ho2021classifier} for both style and content conditioning. Inspired by CFG settings for multiple conditioning in \cite{brooks2023instructpix2pix}, during training, the model observes three conditioning states: full conditioning \((C,z_{ref})\), content-only conditioning \((C,\varnothing)\), and unconditional input \((\varnothing,\varnothing)\). The latter two states are obtained by randomly masking the reference style embedding \(z_{ref}\) or masking both the content and reference embeddings \(C\) and \(z_{ref}\). The score estimate is: 
\begin{equation}
\begin{aligned}
\tilde{s}_{\theta}
={}&
s_{\theta}
\left(
\tilde{Z}_{\sigma}, \sigma;
\varnothing, \varnothing
\right)
\\
&+
w_{C}
\Big[
s_{\theta}
\left(
\tilde{Z}_{\sigma}, \sigma;
C, \varnothing
\right)
-
s_{\theta}
\left(
\tilde{Z}_{\sigma}, \sigma;
\varnothing, \varnothing
\right)
\Big]
\\
&+
w_{ref}
\Big[
s_{\theta}
\left(
\tilde{Z}_{\sigma}, \sigma;
C, z_{ref}
\right)
-
\mathbf{s}_{\theta}
\left(
\tilde{Z}_{\sigma}, \sigma;
C, \varnothing
\right)
\Big]
\end{aligned}
\label{eq:cfg}
\end{equation}
where \(w_{C}\) and \(w_{ref}\) are the guidance strengths for the two conditioning respectively.

We utilize a pretrained audio effects encoder, FxEncoder++ \cite{yeh2025fx}, to extract the content-invariant audio effects features for \(Z\) and \(z_{ref}\). Following MEGAMI \cite{moliner2026automatic}, the content encoder combines CLAP \cite{wu2023large} with a wet-to-dry domain adaptor. The adaptor aligns the CLAP embeddings of dry tracks and their processed versions, enabling effects-invariant content features to be extracted solely from wet audio without requiring real wet-dry pairs.

\subsection{Audio Effects Parameter Decoder}
At this stage, our purpose is to translate the audio effects embeddings into audio effects parameters for each track and then pass them to the mixing console to carry out the mixing task. To accomplish this, we need to train a model to execute the audio effects parameter estimation task. Previous research \cite{barkan2019inversynth,barkan2019deep} regards this task as directly optimizing the parameter loss by applying audio effects with random parameters. However, this is a suboptimal objective since different parameter configurations may yield the same sounds, and it is challenging to generalize to out-of-domain data.  DDSP enables a deep learning system to directly optimize the spectral loss \cite{masuda2021synthesizer,yang2023white}, and some recent studies \cite{peladeau2025audio,hayes2025audio} model the probability distribution of the parameter configuration instead of directly predicting the parameters. We compare these two methods to build our decoder.

\subsubsection{Explicit Parameter Estimation}
The first method is to optimize the network directly using spectral loss. We augment the single dry track \(y_{dry}\) to the wet track \(y_{wet}\)  by applying the same audio effects with random parameters in the mixing console as the input to the audio effects encoder. Subsequently, the embedding output from the encoder is passed through a 3-layer Multilayer perceptron (MLP) and then a sigmoid layer to output the audio effects parameters. Next, the differentiable mixing console transforms \(y_{dry}\) into the predicted wet track \(\hat{y}_{wet}\), and the loss is calculated between \(y_{wet}\) and \(\hat{y}_{wet}\). We use the stereo version of multi-resolution STFT (MSS) loss  \cite{steinmetz2021automatic} and the audio effects embedding loss \cite{moliner2026automatic} for training:
\begin{equation}\label{loss}
\mathcal{L_{\text{spec}}}=\mathcal{L}{\text{MSS}_{l+r}}+\mathcal{L}{\text{MSS}_{l-r}}+\mathcal{L}{\text{FXenc}}
\end{equation}
where \(\mathcal{L}{\text{MSS}_{l+r}}\) represents the loss of the sum of the left and right channel, \(\mathcal{L}{\text{MSS}_{l-r}}\) represents loss of the difference between the two channels, and \(\mathcal{L}{\text{FXenc}}\) is the cosine distance between the embeddings from the audio encoder. We denote the model using this decoder as Diff2Mix-P.

\subsubsection{Modelling Distribution in Parameter Space}
Instead of directly predicting the parameters, we extend the idea of a parameter mode distribution modelling method \cite{peladeau2025audio} to construct our decoder for the audio effects chain. Given the track-level audio effects embedding \(z_i\),  it is first encoded into the mean \(\mu\) and variance \(\sigma\) of a multivariate Gaussian distribution \(p_{\theta}(z_{0}|z_i)\) via a 2-layer MLP. Then, \(z_{0}\) is sampled from this distribution and passed through a 2-layer deep sigmoidal flow (DSF) \cite{huang2018neural} and a sigmoid layer to output the audio effects parameters. Compared to the loss to train Diff2Mix-P, a regularization entropy term is added to balance the estimation accuracy and diversity:
\begin{gather}
\mathcal{L}=\mathcal{L_{\text{spec}}}-\beta\mathcal{H}\!\left(p_{\theta}(z_{0}|z_i)\right)\\
\mathcal{H}\!\left(p_{\theta}(z_{0}|z_i)\right)
= \frac{1}{2}\ln \lvert \det (\sigma I) \rvert + \frac{N}{2}\bigl(1+\ln(2\pi)\bigr)
\end{gather}
where \(\beta\) is a weighting coefficient, which is set to 0.1 in our experiments, and \(N\) is the number of audio effects parameters. More derivation details can be found in \cite{peladeau2025audio}. During the training stage, \(z_{0}\) is randomly sampled. At the inference stage for evaluation, we select the most probable point \(\mu\) of the distribution to mitigate the influence of randomness in the decoder. It should be noted that the mean of the Gaussian distribution does not guarantee the best parameter combination perceptually. We denote the model using this decoder as Diff2Mix-D.

\begin{figure}[t]
  
  \centering
  \includegraphics[width=0.8\columnwidth]{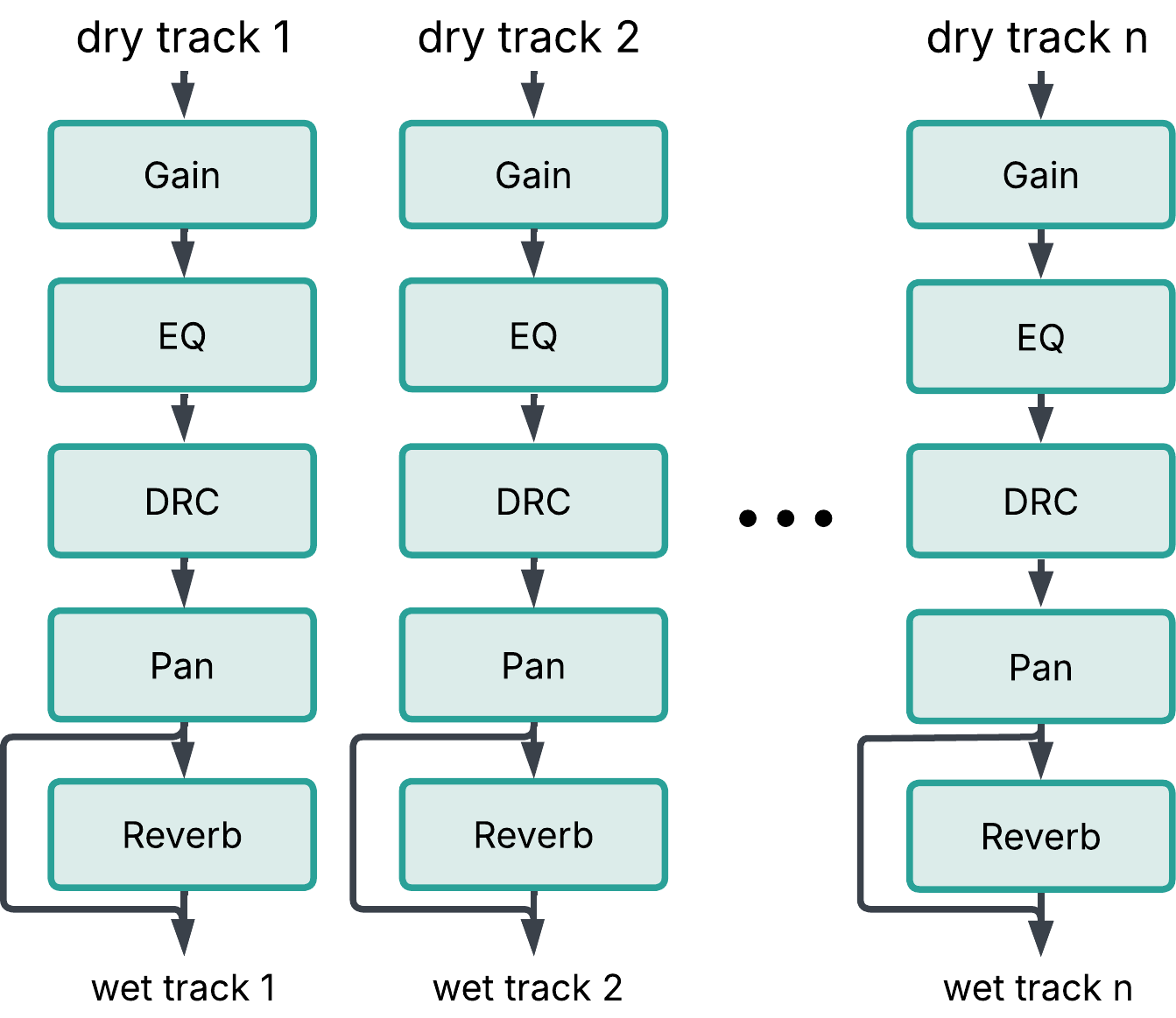}
  \caption{Differentiable Mixing Console}
  \label{console}
  
\end{figure}   
\subsubsection{Mixing Console}
Based on the design of the parameter decoder, our mixing console has an identical fixed audio effects chain for each track. All audio effect processors are differentiable to enable gradient backpropagation during training, and we adopt the implementation in dasp-pytorch\footnote{https://github.com/csteinmetz1/dasp-pytorch/}. As shown in Figure \ref{console}, the audio effects chain is structured as Gain$\rightarrow$ Parametric Equalizer (EQ)$\rightarrow$ Dynamic Range Compressor (DRC)$\rightarrow$ Stereo Panning$\rightarrow$ Artificial Reverberation (Reverb). Reverb is applied as a send effect; that is, the Reverb processes a copy of the input signal and mixes it back with the input signal. This allows the network to learn the appropriate amount of reverberation by controlling the added wet signal \cite{marco_a_martinez_ramirez_2022_7316688}.

\section{Experiments}\label{sec:experiments}
\subsection{Data}
 We curated 196 multitracks from MedleyDB \cite{bittner2014medleydb,bittner2016medleydb} and 240 multitracks from MoisesDB \cite{pereira2023moisesdb}, of which 90\% were allocated for diffusion model training, and the remaining 10\% were for objective evaluation. For the training of the parameter estimation model, similar to MEGAMI, we used multiple datasets including MedleyDB \cite{bittner2014medleydb,bittner2016medleydb}, OpenSinger \cite{huang2021multi}, IDMT-SMT Drums \cite{dittmar2014real}/Bass\cite{abesser2010feature}/Guitar\cite{kehling2014automatic} and GuitarSet \cite{xi2018guitarset}. This enables our model to learn how to handle a variety of music materials.
\subsection{Baselines}\label{baseline}
We compared our models against multiple baselines in mixing quality and style controllability.
\subsubsection{Mixing Quality Baselines}\label{baseline1}
For automatic mixing quality, we compared against 3 baselines.
\begin{itemize}
\item \textbf{Equal Loudness}: A perceptual loudness balancing method.
\item \textbf{FxNorm-automix} \cite{marco_a_martinez_ramirez_2022_7316688}: a supervised deep learning baseline trained with effect-normalized out-of-domain data. We evaluated the best-performing \textit{Ours-S-Lb} version of this model.
\item \textbf{MEGAMI} \cite{moliner2026automatic}: a generative automatic mixing framework operating in an effect-embedding space, which maintains the state-of-the-art performance in automatic mixing. We evaluated the public version that is trained on the same datasets as ours.
\end{itemize}
\subsubsection{Style Transfer Baselines}\label{baseline2}
For style controllability, we compared against two baselines.
\begin{itemize}
\item \textbf{Diff-MST} \cite{soumya_sai_vanka_2024_14877399}: a reference-guided multitrack mixing style transfer system with a differentiable mixing console. We retrained the best-performing model \textit{AF-16} on MedleyDB and multitrack data from Cambridge Music Technology (Cambridge-mt)\footnote{https://cambridge-mt.com/}.
\item \textbf{ST-ITO} \cite{christian_j_steinmetz_2024_14877423}: an inference-time optimization method for audio production style transfer over effect parameters. We evaluated the model to control the VST version of the audio effects chain, which includes distortion, EQ, DRC, feedback delay, and Reverb. Note that ST-ITO directly conducts the style transfer on the mixes rather than the multitrack input. Therefore, to use this method, we employed equal-loudness mixes as its input. This might have a strong impact on the output quality since the model is designed for well-produced songs.
\end{itemize}

\subsection{Evaluation Metrics}\label{metric}
For evaluating the audio mixing quality, we selected Kernel Audio Distance (KAD) \cite{chung2025kad} as in MEGAMI. KAD is a distributional metric based on Maximum Mean Discrepancy between embeddings. We calculated KAD between model outputs and human mixes using several models that produce audio effects embeddings: AFxRep \cite{christian_j_steinmetz_2024_14877423}, FxEncoder \cite{koo2023music}, and FxEncoder++ \cite{yeh2025fx}.

For evaluating the style control ability, we report high-level audio feature loss associated with audio production styles, including dynamics, spatialization and spectral attributes. These features are spectral centroid error (SCE), RMS energy error (RMS), perceptual loudness error (LUFS), mel-spectral distance (MSD), stereo width error (SW) and stereo imbalance error (SI).

\section{Objective Evaluation}

\begin{table}[t]

\begin{center}
\resizebox{\columnwidth}{!}{%
\begin{tabular}{c|ccc}
\toprule
 & AFxRep & FxEncoder & FxEncoder++ \\
\midrule
Equal Loudness & 40.40 & 45.33 & 31.15 \\
FxNorm-automix & 21.83 & 33.69 & 23.99  \\
MEGAMI & \textbf{18.82} & 33.18 & \textbf{12.33}  \\
Diff2Mix-P & 29.21 & 41.69 & 15.45 \\
Diff2Mix-D & 29.27 & \textbf{30.60} & 15.83  \\
\bottomrule
\end{tabular}%
}

\end{center}
\caption{Kernel Audio Distance}
\label{mixquality_}
\end{table}

\begin{table}[t]

\begin{center}
\resizebox{\columnwidth}{!}{%
\begin{tabular}{c|cccccc}
\toprule
 & SCE↓ & RMS↓ & LUFS↓ & MSD↓ & SW↓ & SI↓ \\
\midrule

Diff-MST & 1490.00 & 5.28 & 3.18 & 1.37 & 3.34 & \textbf{0.21} \\
ST-ITO & 1393.15 & 5.75 & 5.24 & 1.28 & 13.04 &  0.57 \\
Diff2Mix-P & 1400.93 & \textbf{3.90} & 3.10 & 1.12 & \textbf{1.79} & 0.63 \\
Diff2Mix-D & \textbf{1288.95} & 4.18 & \textbf{2.22} & \textbf{1.10} & 1.86 & 0.61 \\
\bottomrule
\end{tabular}%
}

\end{center}

\caption{Style Transfer Audio Features Difference}
\label{stylecontrol}
\end{table}

\begin{figure*}[!t]
    \centering
    \includegraphics[width=\textwidth]{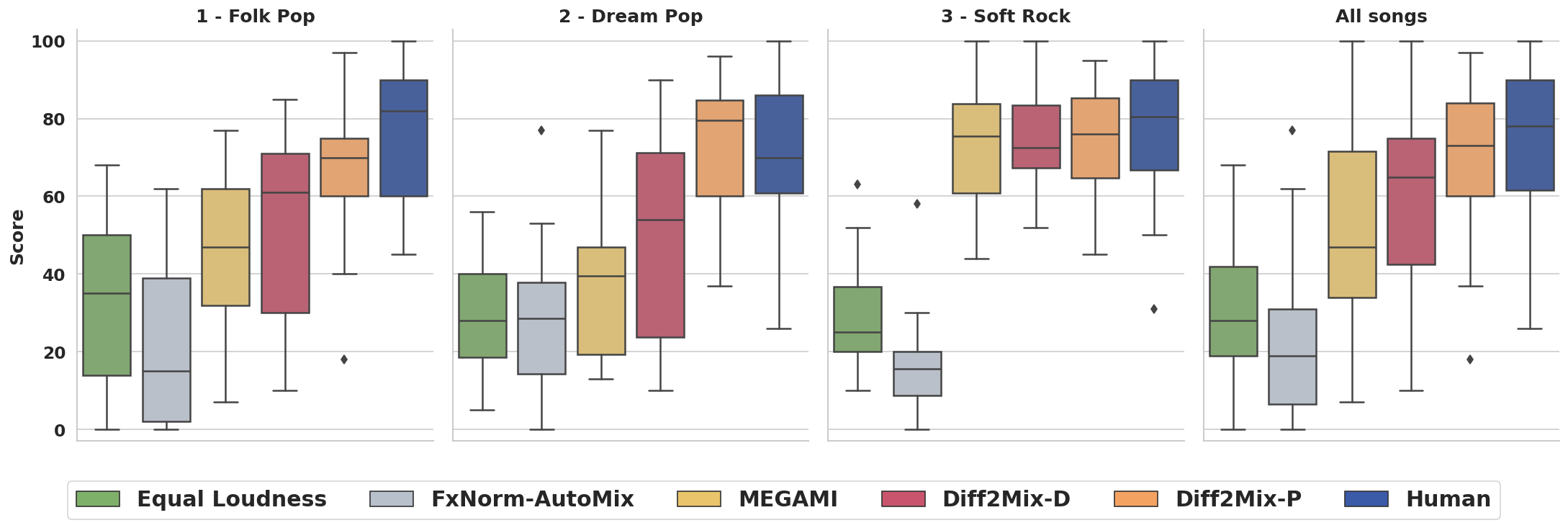}
    \caption{Listening test results of automatic mixing quality rating. The first three plots are ratings corresponding to each multitrack, and the plot on the right side shows the overall ratings.}
    \label{automixplot}
\end{figure*}

\begin{figure*}[!t]
    \centering
    \includegraphics[width=\textwidth]{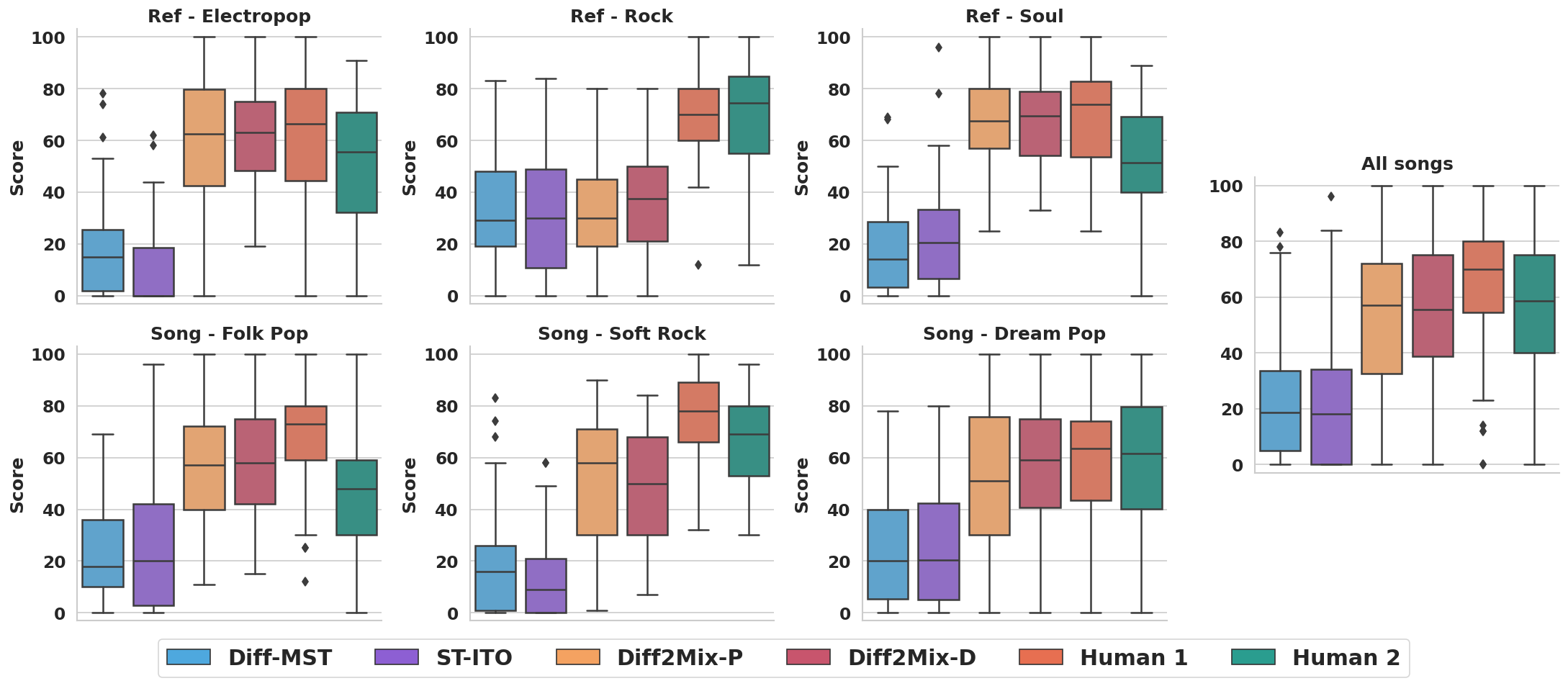}
    \caption{Listening test results of style similarity rating. The three plots in the first row are ratings corresponding to each reference song, and the three plots in the second row are ratings corresponding to each multitrack. The plot on the right side shows the overall ratings.}
    \label{styletransferplot}
\end{figure*}
\subsection{Mixing Quality}
To evaluate the automatic mixing quality, we inferred our Diff2Mix-P and Diff2Mix-D without the input of a reference song. In practice, we masked the global mix audio effects embedding and concatenated it with the content embedding. We compared the KAD scores of our proposed models and three baselines mentioned in Section \ref{baseline1}. Table \ref{mixquality_} summarizes the objective mixing quality results measured by KAD, where lower values indicate that the generated mix distribution is closer to the reference distribution in the corresponding embedding space. Note that FxEncoder++ embedding is used to train MEGAMI and our models, so it may not fairly reflect the model performance. Our models and MEGAMI generally exhibit more competitive performance compared to equal loudness and FxNorm-automix. MEGAMI achieves the best performance among AFxRep and FxEncoder++ representations, while Diff2Mix-D achieves the best in FxEncoder. This is consistent with expectations since MEGAMI translates the audio effects embeddings to wet audio through a black-box model, which is free from the restrictions of fixed audio effects and chain order. However, black-box methods cannot provide the same level of interpretability and control ability as the DDSP method, which significantly limits their application in practical work. Compared to Diff2Mix-D, Diff2Mix-P achieves slightly better results in AFxRep and FxEncoder++, but it performs worse in FxEncoder. We selected the mean of the base distribution to infer Diff2Mix-D, so that potentially sampling better results can be achieved.

\subsection{Style Control}

To evaluate the production style control ability, we conduct a style transfer task and calculate the audio feature difference metrics listed in Section \ref{metric} between model outputs and reference songs. We select three songs from Cambridge-mt as our references.
Table~\ref{stylecontrol} reports the style transfer feature loss. Overall, our methods show the best performance under these metrics. Diff2Mix-D achieves the best performance on SCE, LUFS, and MSD, while remaining close to the best model on SW. Diff2Mix-P performs optimally on RMS and SW, potentially indicating that direct parameter estimation can better preserve some energy and spatial-related attributes. In contrast, the baseline methods are only competitive on SI. Overall, both of the proposed variants demonstrate certain advantages, with Diff2Mix-D showing stronger consistency in overall production style, while Diff2Mix-P also has capabilities very close to those of the former. 
\begin{table}[!t]
\centering

\resizebox{\columnwidth}{!}{%
\begin{tabular}{lcccccc}
\toprule
Model & Equal Loudness & FxNorm-AutoMix & MEGAMI & Diff2Mix-D & Diff2Mix-P & Human \\
\midrule
Equal Loudness  & ---         & $< 0.05$  & $< 0.01$  & $< 0.001$ & $< 0.001$ & $< 0.001$ \\
FxNorm-AutoMix & $< 0.05$    & ---        & $< 0.01$  & $< 0.001$ & $< 0.001$ & $< 0.001$ \\
MEGAMI         & $< 0.01$    & $< 0.01$   & ---        & $< 0.05$  & $< 0.01$  & $< 0.01$  \\
Diff2Mix-D     & $< 0.001$   & $< 0.001$  & $< 0.05$   & ---        & $< 0.05$  & $< 0.01$  \\
Diff2Mix-P     & $< 0.001$   & $< 0.001$  & $< 0.01$   & $< 0.05$  & ---        & \textcolor{red}{$> 0.05$}  \\
Human          & $< 0.001$   & $< 0.001$  & $< 0.01$   & $< 0.01$  & \textcolor{red}{$> 0.05$}  & ---        \\
\bottomrule
\end{tabular}%
}
\caption{Holm-adjusted p-values for pairwise Wilcoxon signed-rank post hoc comparisons for all songs' subjective automatic mixing quality ratings. Red text indicates the difference between these two methods is not statistically significant.}
\label{subjective_pairwise}
\end{table}
\begin{table}[!t]
\centering

\resizebox{\columnwidth}{!}{%
\begin{tabular}{lcccccc}
\toprule
Model & Diff-MST & ST-ITO & Diff2Mix-P & Diff2Mix-D & Human 1 & Human 2 \\
\midrule
Diff-MST   & ---        & \textcolor{red}{$> 0.05$}  & $< 0.001$ & $< 0.001$ & $< 0.001$ & $< 0.001$ \\
ST-ITO     & \textcolor{red}{$> 0.05$}   & ---        & $< 0.001$ & $< 0.001$ & $< 0.001$ & $< 0.001$ \\
Diff2Mix-P & $< 0.001$  & $< 0.001$  & ---        & \textcolor{red}{$> 0.05$}  & $< 0.001$ & \textcolor{red}{$> 0.05$}  \\
Diff2Mix-D & $< 0.001$  & $< 0.001$  & \textcolor{red}{$> 0.05$}  & ---        & $< 0.01$  & \textcolor{red}{$> 0.05$}  \\
Human 1    & $< 0.001$  & $< 0.001$  & $< 0.001$ & $< 0.01$  & ---        & $< 0.05$  \\
Human 2    & $< 0.001$  & $< 0.001$  & \textcolor{red}{$> 0.05$}  & \textcolor{red}{$> 0.05$}  & $< 0.05$  & ---        \\
\bottomrule
\end{tabular}%
}
\caption{Holm-adjusted p-values for pairwise Wilcoxon signed-rank post hoc comparisons for all songs' subjective style similarity ratings. Red text indicates the difference between these two methods is not statistically significant.}
\label{tab:style_similarity_pairwise}
\end{table}
\section{Subjective Evaluation}
We conducted a listening test to compare the performance of the models. We selected three multitrack and three reference songs from different musical genres in the Cambridge-mt dataset. Then, we formulated three questions regarding automatic mixing quality for each multitrack song and nine questions about production style similarity for each multitrack-reference pair, resulting in a total of 12 questions. 
For the mixing quality questions, we compared our models, the models in Section \ref{baseline1}, and professional level human mixes of the multitracks. For the style transfer questions, instead of using our models and the baselines in Section \ref{baseline2}, we recruited two mix engineers to mix the multitracks, with the mixing style being based on the provided reference. 
We asked the participants to rate the mixing quality and the style similarity to the reference on a scale from 0 to 100. Additionally, we instructed them to rate the best one above 80 and the worst below 20 to fully utilize the scale. The listening test was carried out online using webMUSHRA \cite{schoeffler2018webmushra}. In total, we obtained ratings from 20 valid participants who had professional audio mixing experience.

Figure~\ref{automixplot} shows per multitrack and overall subjective ratings for automatic mixing quality; the corresponding statistical test results are shown in Table~\ref{subjective_pairwise}. Overall, the proposed systems receive consistently stronger listener preference compared with equal loudness and FxNorm-automix. Diff2Mix-P reaches the level of human mixing and is likely to be overall better than Diff2Mix-D (\(0.01<p<0.05\)), and Diff2Mix-D is also likely to be better than MEGAMI (\(0.01<p<0.05\)). In the case of per multitrack analysis, Diff2Mix-P, Diff2Mix-D and MEGAMI all perform at the human level of the song in the soft rock genre. However, only Diff2Mix-P maintains this level in the dream pop song, and it slightly outperforms Diff2Mix-D and MEGAMI in the folk pop song.  Our objective evaluation suggests MEGAMI is overall better than the proposed methods, which is consistent with our experience in our pilot listening study, where MEGAMI is more stable in some cases. Due to the limitation of listening tests, in which only three multitracks were included, MEGAMI might be underrated because of the low ratings in the dream pop song case.

Figure~\ref{styletransferplot} reports the style similarity ratings in relation to the reference songs, with statistical test results shown in Table~\ref{tab:style_similarity_pairwise}. 
Overall, our proposed methods achieve human-level performance and outperform the baselines. Diff2Mix-P, Diff2Mix-D and Human 2 show similar performance, and they all perform slightly worse than Human 1. In terms of each multitrack, the models' performance is consistent with the overall trend. In terms of each reference, the proposed models still perform consistently with the overall results for the electropop and soul references, but they all fail in the case of the rock song.  A possible explanation is that the reference song is represented by a single global style embedding, while some rock-specific mixing characteristics depend on track-specific and inter-track interactions that are more difficult to capture with this shared conditioning signal.

\section{Conclusion}
In this paper, we presented Diff2Mix, a controllable automatic mixing framework that integrates a reference-conditioned diffusion model in the track-level audio effects embedding space with a differentiable mixing console. The framework supports user control at two levels: high-level guidance of the overall production style via a reference song and low-level adjustment through explicit audio effects parameters. To convert generated embeddings into effect controls, we investigated both direct parameter estimation and distribution-based parameter modelling. Objective evaluations demonstrate competitive mixing quality and strong performance on production style related metrics, while subjective listening tests indicate that Diff2Mix produces perceptually convincing mixes and effectively transfers production style from a reference song. These findings highlight the potential of combining diffusion-based, style-conditioned generation with differentiable audio processing for controllable intelligent music production. Future work will extend the framework to more diverse or dynamic effect chains, and introduce more user control, such as a text control interface. A lightweight, low-latency implementation could be another promising research direction for real-time mixing.

\section{Acknowledgments}
This work was partly supported by the Yamaha / RAEng Research Chair in Audio Engineering, 2026-2031.

\bibliography{ISMIRtemplate}

%
%
%
%

\end{document}